\documentclass[aip,jcp,preprint,showpacs,superscriptaddress,groupedaddress]{revtex4-1}

\usepackage{amsmath}
\usepackage{amssymb}
\usepackage{braket}
\usepackage{bm}

\begin{document}

\title{Reduced density matrices of Richardson-Gaudin states in the Gaudin algebra basis}
\author{Charles-\'{E}mile Fecteau}
\author{Hubert Fortin}
\author{Samuel Cloutier}
\author{Paul A. Johnson}
 \email{paul.johnson@chm.ulaval.ca}
 
 \affiliation{D\'{e}partement de chimie, Universit\'{e} Laval, Qu\'{e}bec, Qu\'{e}bec, G1V 0A6, Canada}

\date{\today}

\begin{abstract}
Eigenvectors of the reduced Bardeen-Cooper-Schrieffer Hamiltonian have recently been employed as a variational wavefunction ansatz in quantum chemistry. This wavefunction is a mean-field of pairs of electrons (geminals). In this contribution we report optimal expressions for their reduced density matrices in both the original physical basis and the basis of the Richardson-Gaudin pairs. Physical basis expressions were originally reported by Gorohovsky and Bettelheim\cite{GB:2011}. In each case, the expressions scale like $\mathcal{O}(N^4)$, with the most expensive step the solution of linear equations. Analytic gradients are also reported in the physical basis. These expressions are an important step towards practical mean-field methods to treat strongly-correlated electrons.
\end{abstract}

\maketitle

\section{Introduction}

Accurate and affordable treatment of strongly-correlated electrons remains a problem in quantum chemistry. In these systems, many Slater determinants are required to capture the correct physical behaviour. If the number of important Slater determinants is small enough, active space methods are affordable and effective. However, as the number of important Slater determinants increases, this becomes impractical and other avenues must be explored. A promising route is to consider wavefunctions composed of weakly-interacting pairs of electrons (geminals).  This idea dates to the origins of quantum chemistry,\cite{hurley:1953,silver:1969,silver:1970} though has recently been quite fruitful.\cite{coleman:1997,surjan:1999,surjan:2012,neuscamman:2012,peter:2013,stein:2014,boguslawski:2014a,boguslawski:2014b,boguslawski:2014c,tecmer:2014,henderson:2014a,henderson:2014b,shepherd:2014,bulik:2015,kobayashi:2010}

Henderson, Scuseria, and their co-authors are developing a mean-field theory built upon the antisymmetrized geminal power (AGP).\cite{henderson:2019} In particular, they have found an effective algorithm for evaluating the necessary reduced density matrix (RDM) elements,\cite{khamoshi:2019} strategies to include linearly-independent excitations along with their AGP mean-field,\cite{dutta:2020} methods to add dynamic correlation,\cite{henderson:2020a}, computed properties at finite temperature\cite{harsha:2020} and have employed it on a quantum computer.\cite{khamoshi:2020} They have applied their model to the reduced Bardeen-Cooper-Schrieffer\cite{bardeen:1957a,bardeen:1957b} (BCS) pairing model. Strictly speaking, AGP is \emph{not} an eigenvector of the reduced BCS Hamiltonian, though in the large pairing strength limit AGP approaches the exact solution asymptotically. In similar systems, specifically XXZ Richardson-Gaudin (RG) models, pair condensation does occur.\cite{stijn:2014}

Recently,\cite{johnson:2020} we reported the use of the eigenvectors of the reduced BCS Hamiltonian, the RG states,\cite{richardson:1963,richardson:1964,richardson:1965,gaudin:1976} as a mean-field wavefunction ansatz to describe strong electron correlation. That contribution was a first step in the development of many-body methods built upon RG states. RG states are also being employed in nuclear structure\cite{stijn:2017} and condensed matter theory.\cite{claeys:2017a}  Our results were promising, though we explicitly noted many issues to be addressed in upcoming contributions. In particular the energy functional was not optimal. A practical numerical procedure generally requires a solution to three separate problems: i) a method to generate a good initial guess, ii) a cheap method to evaluate the objective function, and iii) an effective numerical solver. In this contribution we aim to completely solve the second problem. We report computationally cheaper expressions for the 1- and 2-body reduced density matrix (RDM) elements both in the original physical basis and the basis of the RG pairs. The first picture is analogous to atomic orbitals (AO) in quantum chemistry, while the second corresponds to the molecular orbital (MO) basis. We will refer to the former as the \emph{physical basis} (PB) and the latter as the \emph{Gaudin basis} (GB).

The optimal RDM formulae in the PB were computed by Gorohovsky and Bettelheim \cite{GB:2011}, so we develop them briefly before adding an expression for the analytic gradient of the RG energy functional.
We have computed the RDMs in the GB for two reasons. First, they may be more effective to evaluate numerically based on the size of the system. Second, to develop perturbation theories, it is often more convenient to work in the GB. Transition density matrices will be reported along with numerical tests in a following contribution. It is worth highlighting that the RG wavefunction is not just a wavefunction ansatz, but an eigenvector of a model Hamiltonian. Thus we are working with not one wavefunction, but a complete set from which we can construct perturbation theories and Green's functions.

The RDM elements are not complicated. The development is tricky and tedious but the final results are simple. All of the final expressions are computed from solutions of sets of linear equations, even sharing the same matrix. Linear equations are numerically very easy to solve, and we need only take simple sums of the results.

In the next section we summarize as briefly as possible the relevant results concerning RG eigenvectors: they are eigenvectors of the reduced BCS Hamiltonian which may itself be written as a linear combination of mutually commuting objects. In section \ref{sec:aobasis} we develop the optimal expressions for the RDMs in the physical basis. In addition, we develop the analytic gradient for the energy functional \eqref{eq:AO_energy}. In section \ref{sec:mobasis} we develop the RDMs in the basis of RG pairs. Readers only interested in the final results are directed to section \ref{sec:summary} where the final RDM expressions are presented as cleanly as possible.

\section{RG Eigenvectors}
In this section we summarize all the relevant results concerning the RG eigenvectors, first from the perspective of a specific Hamiltonian $\hat{H}_{BCS}$, then from the more general perspective of a generating function $S^2(u)$ of conserved quantities. Both approaches lead to the same eigenvectors. The first approach is a specific physical model defined by a set of single-particle energies $\{\varepsilon\}$ and a pairing strength $g$, while the second approach relies only on the algebraic structure of the pairs of electrons. For a general overview of the first approach see refs\cite{dukelsky:2004,ortiz:2005} and for the second approach refs.\cite{sklyanin:1989,faribault:2014} 

Before proceeding to the solutions we will briefly denote our conventions. The fundamental objects are a pairing representation of su(2):

\begin{align} \label{eq:su2}
S^+_i = a^{\dagger}_{i\uparrow} a^{\dagger}_{i\downarrow}, \quad S^-_i = a_{i\downarrow} a_{\uparrow}, \quad S^z_i = \frac{1}{2} \left( a^{\dagger}_{i\uparrow} a_{i\uparrow} + a^{\dagger}_{i\downarrow}a_{i\downarrow} -1 \right) 
\end{align}
in which $a^{\dagger}_{i\uparrow}$ creates an up spin electron in spatial orbital $i$, $a_{i\downarrow}$ removes a down spin electron etc. It will often be convenient to use the number operator $\hat{n}_i$ rather $S^z_i$, as the two are simply related:
\begin{align}
\hat{n}_i = 2 S^z_i + 1.
\end{align}

With a complex number $u$, define the pair creators:
\begin{align}
S^+(u) = \sum_i \frac{S^+_i}{u-\varepsilon_i}.
\end{align}
The RG eigenvectors are products
\begin{align} \label{eq:RGvec}
\ket{\{v\}} = S^+(v_1) S^+(v_2) \dots S^+(v_M) \ket{\theta}
\end{align}
where the complex numbers $\{v\}$, which we call rapidities, are solutions of Richardson's equations, 
\begin{align} \label{eq:Rich}
\lambda_a = \frac{2}{g} + \sum_i \frac{1}{v_a - \varepsilon_i} + \sum_{b\neq a} \frac{2}{v_b - v_a} = 0.
\end{align}
Each $\lambda_a$ must vanish numerically. Distinct eigenvectors correspond to distinct sets of rapidities. Unlike the case for electrons \cite{laurie}, none of the rapidities coincide for different eigenvectors. In equation \eqref{eq:RGvec}, $\ket{\theta}$ is a vacuum with respect to all $S^-_i$. Usually it is the empty state, but can also include unpaired (non-interacting) electrons.

There are $N$ spatial orbitals in which to place pairs. Labels corresponding to the spatial orbitals are labelled with indices $i,j,k,l$ etc. There are $M$ pairs of electrons. Labels corresponding to the pairs are labelled with indices $a,b,c,d$ etc. The set $\{v\}$ will always denote a solution of Richardson's equations (here the ground state), while $\{u\}$ denotes an arbitrary set of complex numbers. 
\subsection{Reduced BCS Hamiltonian}
The reduced BCS Hamiltonian expresses competition between a filling of the lowest single-particle states $\{\varepsilon\}$ and a constant pairing interaction $g$:
\begin{align} \label{eq:BCSham}
\hat{H}_{BCS} = \frac{1}{2} \sum_i \varepsilon_i \hat{n}_i -\frac{g}{2} \sum_{ij} S^+_iS^-_j
\end{align}
To show that the RG states \eqref{eq:RGvec} are eigenvectors of \eqref{eq:BCSham}, one strategy is to move $\hat{H}_{BCS}$ past each $S^+(v)$ and collect terms. We must therefore evaluate the following single-
\begin{align}
[\hat{H}_{BCS},S^+(v_1)] = v_1 S^+(v_1) - \sum_i S^+_i +g \sum_{ij} \frac{S^+_i S^z_j}{v_1-\varepsilon_j}
\end{align}
and double-commutators 
\begin{align}
[[\hat{H}_{BCS},S^+(v_1)],S^+(v_2)] = \frac{g}{v_2-v_1} \sum_i S^+_i \left( S^+(v_1) - S^+(v_2) \right).
\end{align}
Now, we can collect terms
\begin{align}
\hat{H}_{BCS} \ket{ \{v\}} &= \sum_a \prod_{b\neq a} S^+(v_b) [ \hat{H}_{BCS},S^+(v_a) ] \ket{\theta} \nonumber \\
&+ \sum_{a \neq b} [ [ \hat{H}_{BCS}, S^+(v_a) ],S^+(v_b) ] \prod_{c\neq a,b} S^+(v_c) \ket{\theta} \\
&= \sum_a v_a \ket{\{v\}} - \frac{g}{2} \sum_i S^+_i \sum_a \lambda_a \prod_{b \neq a} S^+(v_b) \ket{\theta}.
\end{align}
In the last line, there is one term proportional to \eqref{eq:RGvec} and a collection of unwanted terms proportional to Richardson's equations \eqref{eq:Rich}. Thus, the RG state is an eigenvector of $\hat{H}_{BCS}$ provided that Richardson's equations are satisfied.

\subsection{Transfer matrix}
In addition to the pair creator $S^+(u)$, the Gaudin algebra has two more objects
\begin{align} \label{eq:galg}
S^-(u) = \sum_i \frac{S^-_i}{u-\varepsilon_i}, \quad S^z(u) = \frac{1}{g} - \sum_i \frac{S^z_i}{u-\varepsilon_i}
\end{align}
that have the structure:
\begin{subequations}
\begin{align}
[S^+(u),S^-(v) ] &= 2 \frac{S^z(u) - S^z(v)}{u-v} \\
[S^z(u),S^+(v) ] &= \frac{S^+(u) - S^+(v)}{u-v} \\
[S^z(u),S^-(v) ] &= - \frac{S^-(u) - S^-(v)}{u-v}
\end{align}
\end{subequations}
and the intuitive result
\begin{align}
[S^+(u),S^+(v)] = [S^-(u),S^-(v)] = [S^z(u),S^z(v)]=0.
\end{align}
In the case of repeated arguments, a limiting procedure can be adopted,\cite{ortiz:2005} with $h$ a small, positive real number
\begin{align}
[S^+(u), S^-(u)] = \lim_{h\rightarrow0} [S^+(u+h),S^-(u)] = \lim_{h\rightarrow0} 2 \frac{S^z(u+h)-S^z(u)}{h} = 2 \frac{\partial S^z(u)}{\partial u}
\end{align}
which introduces the derivative of $S^z(u)$ with respect to $u$. 

In the Gaudin algebra approach, the RG states are eigenvectors\cite{sklyanin:1989} of $S^2(u)$:
\begin{align}
S^2(u) &= S^z(u) S^z(u) + \frac{1}{2}\left( S^+(u)S^-(u) + S^-(u)S^+(u) \right) \\
&= S^z(u) S^z(u) + S^+(u)S^-(u) - \frac{\partial S^z(u)}{\partial u}
\end{align}
It should be noted that $S^2(u)$ is not a Casimir operator, meaning that it does not commute with everything like $S^2_i$ does in the local su(2) copies. Rather, we can evaluate its commutators with the pair creators:
\begin{align}
[S^2(u),S^+(v_1)] = \frac{2}{u-v} \left( S^+(u)S^z(v_1) - S^+(v_1)S^z(u) \right)
\end{align}
\begin{align}
[[S^2(u),S^+(v_1)],S^+(v_2)] = \frac{2 S^+(u)S^+(v_1)}{(u-v_2)(v_1-v_2)} + \frac{2 S^+(u)S^+(v_2)}{(u-v_1)(v_2-v_1)} + \frac{2 S^+(v_1)S^+(v_2)}{(u-v_1)(u-v_2)}
\end{align}
to move $S^2(u)$ to the right until it acts on the vacuum.
\begin{align}
S^2(u) \ket{\{v\}} &= \prod_a S^+(v_a) S^2(u) \ket{\theta} + \sum_a \prod_{b \neq a} S^+(v_b) [S^2(u),S^+(v_a)]\ket{\theta} \nonumber \\
&+ \sum_{a \neq b} [[S^2(u),S^+(v_a)],S^+(v_b)] \prod_{c \neq a,b} S^+(v_c) \ket{\theta}
\end{align}
The vacuum is an eigenvector of $S^z(u)$ with eigenvalue:
\begin{align}
S^z(u) \ket{\theta} &= \alpha(u) \ket{\theta} \\
\alpha(u) &= \frac{1}{g} + \frac{1}{2} \sum_i \frac{1}{u-\varepsilon_i}.
\end{align}
With these results, the action of $S^2(u)$ upon an RG state is
\begin{align}
S^2(u) \ket{\{v\}} &= \Lambda(u,\{v\}) \ket{\{v\}} + \sum_a \Lambda_a (u,\{v\}) S^+(u) \ket{\{v\}_a} \\
\Lambda(u,\{v\}) &= \alpha(u)^2 - \frac{\partial \alpha(u)}{\partial u} - 2 \sum_a \frac{\alpha(u) - \alpha(v_a)}{u-v_a} \\
\Lambda_a(u,\{v\}) &= \frac{2}{u-v_a} \left( \alpha(u_a)- \sum_{b\neq a} \frac{1}{v_b-v_a} \right)
\end{align}
which is to say the $\ket{\{v\}}$ is an eigenvector of $S^2(u)$ with eigenvalue $\Lambda(u,\{v\})$ provided the numerical coefficients $\Lambda_a(u,\{v\})$ vanish. This is the case when the bracketed terms on the last line vanish, and one can recognize them as identical to Richardson's equations.

The two approaches have a clear connection. First, it is not difficult to show that for any $u_1,u_2$:
\begin{align}
[S^2(u_1),S^2(u_2)]=0.
\end{align}
$S^2(u)$ is thus a generating function for conserved quantities: for any choice of its argument, we know its eigenvectors. In this specific representation, we can rewrite it as\cite{sklyanin:1989}
\begin{align} \label{eq:tmat}
S^2(u) = \frac{1}{g^2} - \frac{2}{g} \sum_i \frac{\hat{R}_i}{u-\varepsilon_i}  + \sum_i \frac{S^2_i}{(u-\varepsilon_i)^2}.
\end{align}
The first and last terms in eq. \eqref{eq:tmat} are constants. Specifically $S^2_i$ is a Casimir operator for the objects \eqref{eq:su2} and thus acts as a constant on them. The other terms involve the objects
\begin{align}
\hat{R}_i = S^z_i - g \sum_{j\neq i} \frac{S^z_i S^z_j + \frac{1}{2} ( S^+_iS^-_j + S^-_iS^+_j ) }{\varepsilon_i - \varepsilon_j}
\end{align}
which are called ``conserved charges'' or ``integrals of motion''. Again, it is not difficult to show that\cite{camb:1997}
\begin{align}
[\hat{R}_i , \hat{R}_j] = 0, \quad \forall i,j
\end{align}
so that we can diagonalize these objects directly, and any linear combination of them will share their eigenvectors. In particular,
\begin{align}
\sum_i \varepsilon_i \hat{R}_i = \hat{H}_{BCS} + cte
\end{align}
with $cte$ an irrelevant constant.

The RG states are thus eigenvectors of $\hat{H}_{BCS}$ and $S^2(u)$ provided that the rapidities are solutions of Richardson's equations. This is the Bethe ansatz\cite{bethe:1931} construction: the original eigenvalue problem has been reduced to a system of non-linear equations to be solved.

\subsection{Richardson's equations}
Many methods have been proposed and employed to solve Richardson's equations. Approaches include clusterization methods,\cite{rombouts:2004} Heine-Stieltjes correspondences,\cite{guan:2012} stochastic methods,\cite{pogosov:2012} pseudo-deformations of su(2),\cite{stijn:2012} and eigenvalue-based methods.\cite{faribault:2011,claeys:2015} We have opted to employ eigenvalue-based solvers as they robust and straightforward. Solving Richardson's equations is now a solved problem. With the so-called eigenvalue based variables (EBV):
\begin{align} \label{eq:ebv}
U_i = \sum_a \frac{1}{\varepsilon_i - v_a}
\end{align}
Richardson's equations can be shown to be equivalent to:
\begin{align}
0 = U^2_i - \frac{2}{g} U_i - \sum_{j\neq i} \frac{U_j-U_i}{\varepsilon_j - \varepsilon_i}.
\end{align}
These equations are much easier to solve as they don't have any divergences in the denominator: the values of $\{\varepsilon\}$ are fixed whereas for Richardson's equations, the rapidities appear explicitly in the denominator. The solution begins with a good guess for $\{U\}$, from which Newton-Raphson yields rapid convergence. From the values of the EBV, one can use Laguerre's method to find the rapidities. For this construction we have assumed that all $\{\varepsilon\}$ are distinct, and hence that each level may be only occupied by a single pair. To correctly account for degeneracy in $\{\varepsilon\}$, the general problem has already been solved as well.\cite{elaraby:2012}

\section{Reduced Density Matrix Elements: Physical Basis}
\label{sec:aobasis}
Our goal is to use the RG states as a variational ansatz for a Coulomb Hamilonian:
\begin{align}
\hat{H}_C = \sum_{ij} h_{ij} \sum_{\sigma} a^{\dagger}_{i\sigma}a_{j\sigma} + \frac{1}{2} \sum_{ijkl} V_{ijkl} \sum_{\sigma \tau} a^{\dagger}_{i\sigma}a^{\dagger}_{j\tau} a_{l\tau}a_{k\tau}.
\end{align}
Here $\sigma$ and $\tau$ are spin labels, and the molecular integrals have been calculated in an orthonormal spatial orbital basis $\{\phi\}$
\begin{align}
h_{ij} &= \int d\mathbf{r} \phi^*_i (\mathbf{r}) \left( - \frac{1}{2} \nabla^2 - \sum_I \frac{Z_I}{| \mathbf{r} - \mathbf{R}_I |} \right) \phi_j (\mathbf{r}) \\
V_{ijkl} &= \int d\mathbf{r}_1 d\mathbf{r}_2 \frac{\phi^*_i(\mathbf{r}_1)  \phi^*_j(\mathbf{r}_2)  \phi_k(\mathbf{r}_1)  \phi_l(\mathbf{r}_2)  }{| \mathbf{r}_1 - \mathbf{r}_2|}.
\end{align}
RG states lie in the seniority-zero sector, meaning that all electrons remain paired in spatial orbitals. Therefore, terms with seniorities other than zero in the Coulomb Hamiltonian will give no contribution when an expectation value is taken. Thus, only the seniority-zero piece of the Coulomb Hamiltonian,
\begin{align} \label{eq:sen0_ham}
\hat{H}_0 = \sum_i h_{ii} \hat{n}_i + \frac{1}{4} \sum_{ij} W_{ij} \hat{n}_i \hat{n}_j + \sum_{ij} V_{iijj} S^+_i S^-_j
\end{align}
will matter, with
\begin{align}
W_{ij} = \begin{cases}
2V_{ijij} - V_{ijji}, \quad &i \neq j \\
0, \quad &i = j.
\end{cases}
\end{align}
The element $W_{ii}$ is set to zero to avoid double-counting the element $V_{iiii}$. Taking an expectation value with the RG state yields the energy functional:
\begin{align} \label{eq:AO_energy}
E[\{\varepsilon\},g] = 2 \sum_i h_{ii} \gamma_i  + \sum_{ij} W_{ij} D_{ij} + \sum_{ij} V_{iijj} P_{ij} .
\end{align}
which is to be optimized variationally for the parameters $\{\varepsilon\}$ and $g$. To evaluate the energy \eqref{eq:AO_energy} we require the reduced density matrix (RDM) elements
\begin{align}
\gamma_i &= \frac{1}{2} \frac{\braket{\{v\} | \hat{n}_i | \{v\}}}{\braket{\{v\}| \{v\}}} \label{eq:AO_1dm} \\
D_{ij} &= \frac{1}{4} \frac{\braket{\{v\} | \hat{n}_i \hat{n}_j | \{v\}}}{\braket{\{v\}| \{v\}}} \\
P_{ij} &= \frac{\braket{\{v\} | S^+_i S^-_j | \{v\}}}{\braket{\{v\}| \{v\}}}.
\end{align}
The RDM elements are functions of $\{\varepsilon\}$ and $g$ though this dependence will be suppressed to keep the equations clean. The 1-RDM $\gamma$ is diagonal, and with this choice of normalization has entries between zero and one. The only non-zero elements of the 2-RDM are the diagonal correlation function $D_{ij}$ and the pair correlation function $P_{ij}$.

While we have previously employed the ingenious expressions of Faribault et al.,\cite{faribault:2008,faribault:2010} the most efficient expressions are obtained with the approach of Gorohovsky and Bettelheim. \cite{GB:2011} There are three tools required. The first is Slavnov's theorem\cite{Slavnov:1989,Belliard:2019} which expresses the scalar product of two states as the determinant of a matrix
\begin{align}
\braket{ \{v \} | \{u\}} &= \frac{\prod_{a\neq b} (v_a - u_b)}{\prod_{a<b} (u_a-u_b)(v_b-v_a)} \det J \label{eq:Slavnov} \\
J_{ab} &= \frac{v_b -u_b}{v_a-u_b} \left( \sum_i \frac{1}{(v_a-\varepsilon_i)(u_b-\varepsilon_i)} - 2 \sum_{c\neq a} \frac{1}{(v_a-v_c)(u_b-v_c)} \right).
\end{align}
In the expression \eqref{eq:Slavnov}, the set $\{v\}$ are solutions of Richardson's equations, while the set $\{u\}$ is arbitrary. This expression is quite practical, as a determinant may be evaluated with a cost of the cube of the size of the matrix. Slavnov's original paper treated eigenvectors of the six-vertex model, which includes spin-conserving Heisenberg models. The RG version (employed herein) is a specific limit originally obtained by Zhou et al. \cite{Zhou:2002} For a general discussion of scalar products and correlation functions for Bethe ansatz wavefunctions, the reader is referred to ref.\cite{korepin_book} 

By taking the limit $\{u\} \rightarrow \{v\}$, we get the norm of the RG state,
\begin{align} \label{eq:gmat}
\braket{\{v\} | \{v\} } &= \det G \\
G_{ab} &= 
\begin{cases}
\sum_i \frac{1}{(v_a -\varepsilon_i)^2} - \sum_{c\neq a} \frac{2}{(v_a -v_a)^2}, \quad &a=b \\
\frac{2}{(v_a-v_b)^2}, \quad &a\neq b
\end{cases}
\end{align}
where the \emph{Gaudin matrix} $G$ is the Jacobian of Richardson's equations. This expression for the norm was known to Richardson \cite{richardson:1965} and Gaudin \cite{gaudin:1976}, from their original papers. 

The second tool is Cramer's rule, an elementary result from linear algebra: for the system of linear equations
\begin{align}
A \textbf{x}=\textbf{b}
\end{align}
the elements of the vector $\textbf{x}$ of solutions are expressible as ratios of determinants
\begin{align} \label{eq:Cramer}
x_a = \frac{\det A^b_a}{\det A}.
\end{align}
In \eqref{eq:Cramer} the matrix $A^b_a$ is the matrix $A$ with the $a$th column replaced by the RHS $\textbf{b}$. So, the $a$th element of the solution is a ratio of two determinants differing by a single column.

The third tool is a theorem of Jacobi,\cite{vein_book} which states that scaled cofactors can be expressed as a determinant of simple scaled cofactors. Practically, we will only need this result to second order, for which Jacobi's theorem gives:
\begin{align}
\frac{\det A^{cd}_{ab}}{\det A} = \frac{\det A^c_a }{\det A} \frac{\det A^d_b}{\det A} - \frac{\det A^d_a}{\det A}\frac{\det A^c_b}{\det A}
\end{align}
where the matrix $A^{cd}_{ab}$ is the matrix $A$ with the $a$th column replaced by the vector $\textbf{c}$ and the $b$th column replaced with the vector $\textbf{d}$. The RHS is a 2 $\times$ 2 determinant, whose entries are ratios of determinants differing by one column. Remarkably, this result holds to any order, meaning that the ratio of two determinants differing by $k$ columns can be expressed as a $k\times k$ determinant whose entries are ratios of determinants differing by one column.

\subsection{1-electron reduced density matrix}

We follow the form factor approach. First, we evaluate $\braket{\{v\} | S^z_i | \{u\}}$ with $\{v\}$ a solution of Richardson's equations, and $\{u\}$ arbitrary. To do this, move $S^z_i$ to the right, past each $S^+(u)$, until it acts on the vacuum.
\begin{align}
\braket{\{v\} | S^z_i | \{u\}} = \sum_a \frac{1}{u_a - \varepsilon_i} \braket{\{v\} | S^+_i | \{u\}_a} - \frac{1}{2} \braket{\{v\} | \{u\}}
\end{align}
Here, the notation $\{u\}_a$ means the set $\{u\}$ without the element $u_a$. Rather than $S^z_i$, we will employ the number operator $\hat{n}_i$, which counts the number of electrons in level $i$, for which,
\begin{align}
\braket{\{v\} | \hat{n}_i | \{u\}} = \sum_a \frac{2}{u_a - \varepsilon_i} \braket{\{v\} | S^+_i | \{u\}_a}. \label{eq:nff}
\end{align}
The scalar products on the right hand side of \eqref{eq:nff}, called \emph{form factors}, are easily evaluated: the local operators $S^+_i$ are the residues of the RG pair $S^+(u)$ at the simple pole $u = \varepsilon_i$, so 
\begin{align}
S^+_i = \lim_{u \rightarrow \varepsilon_i} (u-\varepsilon_i) S^+(u).
\end{align}
Further, the form factor is the residue of the scalar product 
\begin{align}
\braket{\{v\}| S^+_i | \{u\}_a} = \lim_{u_a \rightarrow \varepsilon_i} (u_a-\varepsilon_i) \braket{\{v\}|\{u\}}
\end{align}
given by Slavnov's theorem at the simple pole $u_a \rightarrow \varepsilon_i$.

Now, setting $\{u\}=\{v\}$, we get
\begin{align} \label{eq:gb_ff1}
\braket{\{v\}| S^+_i| \{v\}_a} = (v_a - \varepsilon_i) \det G^i_a.
\end{align}
The matrix $G^i_a$ is the Gaudin matrix \eqref{eq:gmat} used to calculate the norm, with the $a$th column replaced with the $i$th version of the column
\begin{align} \label{eq:local_RHS}
\textbf{b}_i = 
\begin{pmatrix}
\frac{1}{(v_1 - \varepsilon_i)^2} \\
\frac{1}{(v_2 - \varepsilon_i)^2} \\
\vdots \\
\frac{1}{(v_M - \varepsilon_i)^2} \\
\end{pmatrix}.
\end{align}
With \eqref{eq:AO_1dm}, \eqref{eq:nff} and \eqref{eq:gb_ff1} the normalized 1-RDM becomes:
\begin{align}
\gamma_i & = \sum_a \frac{\det G^i_a}{\det G} \label{eq:1rdm}
\end{align}
From Cramer's rule, the elements of the right hand side are particularly simple to compute. In particular, for the system of linear equations
\begin{align}
G\textbf{x} = \textbf{b}_i \label{eq:linear_equations}
\end{align}
the $a$th entry of the vector $\textbf{x}$ is precisely
\begin{align}
x_a = \frac{\det G^i_a}{\det G}.
\end{align}
Thus, to evaluate the 1-RDM, we solve the system of linear equations \eqref{eq:linear_equations} for each right-hand side $\textbf{b}_i$, and save the solutions. Solving linear equations has a scaling of $\mathcal{O}(M^3)$, and there are $N$ sets of linear equations, so this computation has a scaling of $\mathcal{O}(NM^3)$. With the solutions $\{x\}$ stored, each 1-RDM element is easily computed as a sum.

Physically, the solutions of the linear equations \eqref{eq:linear_equations} have a simple interpretation. Starting from Richardson's equations, perturb one $\varepsilon$ and measure the responses in each $v_a$, 
\begin{align}
\varepsilon_k & \mapsto \varepsilon_k + \delta \varepsilon_k \\
v_a & \mapsto v_a + \delta v_a + \mathcal{O}(\delta v_a^2) 
\end{align}
keeping only the linear terms, without too much difficulty:
\begin{align}
\left( \sum_i \frac{1}{(v_a - \varepsilon_i)^2} - \sum_{b\neq a} \frac{2}{(v_b - v_a)^2}  \right) \frac{\delta v_a}{\delta \varepsilon_k} + \sum_{b \neq a} \frac{2}{(v_b - v_a)^2} \frac{\delta v_b }{\delta \varepsilon_k} = \frac{1}{(v_a - \varepsilon_k)^2}.
\end{align}
We can identify the ratios of first order changes as the partial derivatives, i.e. $\frac{\delta v_a}{\delta \varepsilon_k}=\frac{\partial v_a}{\partial \varepsilon_k}$. Further, as there is one such equation for each $v_a$, taken together, they form a linear system of equations, specifically
\begin{align}
G \frac{\partial \mathbf{v}}{\partial \varepsilon_k} = \textbf{b}_k.
\end{align}
From Cramer's rule, we get directly
\begin{align}
\frac{\det G^k_a}{\det G} = \frac{\partial v_a}{\partial \varepsilon_k}.
\end{align}
Finally, this means that 
\begin{align}
\gamma_i = \sum_a \frac{\partial v_a}{\partial \varepsilon_i}.
\end{align}
The simplicity of this result is highly suggestive that is the optimal expression. In terms of computation, we set up the matrix $G$, solve the systems of linear equations and save the results. 

\subsection{2-electron reduced density matrix: Pair-Correlation Function}
Computing the pair-correlation function follows along the same lines. Start with two distinct sets of rapidities and move $S^-_j$ to the right until it acts on the vacuum. The result is:
\begin{align}
\braket{ \{v\} | S^+_i S^-_j | \{u\} } = \sum_a \frac{\braket{ \{v\} | S^+_i | \{u\}_a}}{u_a - \varepsilon_j} 
- \sum_{a\neq b} \frac{ \braket{ \{v\} | S^+_i S^+_j | \{u\}_{a,b} }  }{(u_a - \varepsilon_j)(u_b - \varepsilon_j)}.
\end{align}
As before, $\{u\}_a$ denotes the set $\{u\}$ without the element $u_a$, while $\{u\}_{a,b}$ means $\{u\}$ without $u_a$ and $u_b$. The first term is evaluated in the same manner as in the previous section. In the second term, notice that both terms in the denominator involve $\varepsilon_j$. The numerator may be evaluated in a similar manner. Take the residue of Slavnov's theorem:
\begin{align}
\braket{ \{v\} | S^+_i S^+_j | \{u\}_{a,b} } = \lim_{u_a \rightarrow \varepsilon_i} \lim_{u_b \rightarrow \varepsilon_j} (u_a - \varepsilon_i) (u_b - \varepsilon_j) \braket{ \{v\} | \{u\}}.
\end{align}
Next, take the limit that $\{u\} \rightarrow \{v\}$, the result of which is:
\begin{align} \label{eq:ppff}
\braket{ \{v\} | S^+_i S^+_j | \{v\}_{a,b} } = \frac{(v_a - \varepsilon_i)(v_b- \varepsilon_i)(v_a - \varepsilon_j)(v_b-\varepsilon_j)}{(\varepsilon_i - \varepsilon_j)(v_b-v_a)} \det G^{ij}_{ab}
\end{align} 
Here, the matrix $G^{ij}_{ab}$ is the Gaudin matrix \eqref{eq:gmat} with the  $a$th column replaced with the $i$th RHS \eqref{eq:local_RHS} and the $b$th column replaced with the $j$th RHS \eqref{eq:local_RHS}. Thus $G^{ij}_{ab}$ is the matrix $G$ with two replaced columns. Jacobi's result is that the determinant of a multiply-substituted matrix scaled by the original determinant is the determinant of the scaled simple substitutions. In this case, this means directly:
\begin{align}
\frac{\det G^{ij}_{ab}}{\det G} &= \frac{\det G^{i}_{a}}{\det G} \frac{\det G^{j}_{b}}{\det G} - \frac{\det G^{j}_{a}}{\det G} \frac{\det G^{i}_{b}}{\det G} \\
&= \frac{\partial v_a}{\partial \varepsilon_i} \frac{\partial v_b}{\partial \varepsilon_j} - \frac{\partial v_a}{\partial \varepsilon_j} \frac{\partial v_b}{\partial \varepsilon_i}
\end{align}
Using this result, the pair-correlation function is:
\begin{align}
P_{ij} = \sum_a \frac{v_a-\varepsilon_i}{v_a - \varepsilon_j} \frac{\partial v_a}{\partial \varepsilon_i} -2 \sum_{a<b} \frac{(v_b-\varepsilon_i)(v_a - \varepsilon_i)}{(\varepsilon_i - \varepsilon_j)(v_b-v_a)}
\left( \frac{\partial v_a}{\partial \varepsilon_i} \frac{\partial v_b}{\partial \varepsilon_j} - \frac{\partial v_a}{\partial \varepsilon_j} \frac{\partial v_b}{\partial \varepsilon_i} \right)
\end{align}
Because of Jacobi's identity, evaluating $P_{ij}$ requires only the solutions of the same sets of linear equations as $\gamma_i$. With these values the pair-correlation function may be evaluated with a cost of $\mathcal{O}(N^2 M^2)$: there is a double summation over $M$ elements, and $N^2$ elements of $P_{ij}$.

\subsection{2-electron reduced density matrix: Diagonal Correlation Function}
Proceeding in the same manner as the previous section, we can write
\begin{align}
D_{ij} = \frac{1}{\det G} \sum_{a \neq b} \frac{\braket{\{v\}| S^+_iS^+_j |\{v\}}}{(v_a-\varepsilon_i)(v_b-\varepsilon_j)}
\end{align}
and with the result \eqref{eq:ppff}, this reduces to an expression we may evaluate with the same cost as $P_{ij}$:
\begin{align}
D_{ij} = \sum_{a<b} \frac{(v_a - \varepsilon_i)(v_b - \varepsilon_j) + (v_a - \varepsilon_j)(v_b - \varepsilon_i)}{(\varepsilon_i - \varepsilon_j)(v_b-v_a)} 
\left( \frac{\partial v_a}{\partial \varepsilon_i} \frac{\partial v_b}{\partial \varepsilon_j} - \frac{\partial v_a}{\partial \varepsilon_j} \frac{\partial v_b}{\partial \varepsilon_i} \right)
\end{align}
\subsection{Sum rules and consistency checks}
In this normalization, the 1-RDM counts the number of pairs in each site. Therefore the trace of the 1-RDM should be the total number of pairs. Likewise, the trace of $D_{ij}$ is the square of the number of pairs.
\begin{align}
\sum_i \gamma_i &= M \\
\sum_{ij} D_{ij} &= M^2 
\end{align}

Partial traces of $D_{ij}$ yield 1-RDM elements, scaled by the number of pairs:
\begin{align}
\sum_j D_{ij} = \frac{1}{M} \gamma_i
\end{align}

As the diagonal of $P_{ij}$ is $\gamma_i$, the energy of the reduced BCS Hamiltonian can be computed with $P_{ij}$ and compared with the exact expression (the sum of the rapidities). This defines a consistency check for $P_{ij}$:
\begin{align}
\sum_{ij} \left( \delta_{ij} \varepsilon_i - \frac{g}{2}\right) P_{ij} = \sum_a v_a.
\end{align}
In our variational calculations we have observed that this consistency check is sometimes violated. This is likely due to Laguerre's method failing to produce correct rapidities from the EBV. In such cases, RDM expressions in terms of the EBV would be more robust. Scalar products are known in terms of EBV as are 1-RDM expressions.\cite{claeys:2017b} Expressions for $P_{ij}$ and $D_{ij}$ are not yet known however.

\subsection{Analytic gradient}
With the same machinery required to compute the reduced density matrices, we may compute the analytic gradient of the energy functional \eqref{eq:AO_energy} with respect to the variational parameters $\{\varepsilon\},g$. Starting with the energy expression \eqref{eq:AO_energy}, we can use the results for the 1- and 2-RDMs to write:
\begin{align}
E &= 2 \sum_{ja} \left( h_{jj} + \frac{1}{2} V_{jjjj}\right)  \frac{\partial v_a}{\partial \varepsilon_j} + \sum_{i\neq j} V_{iijj} \sum_a \frac{v_a- \varepsilon_i}{u_a - \varepsilon_j}\frac{\partial v_a}{\partial \varepsilon_i} 
+ \sum_{\substack{ i\neq j \\ a<b}} T^{ab}_{ij} \left( \frac{\partial v_a}{\partial \varepsilon_i} \frac{\partial v_b}{\partial \varepsilon_j} - \frac{\partial v_a}{\partial \varepsilon_j} \frac{\partial v_b}{\partial \varepsilon_i} \right)
\end{align}
with the intermediate:
\begin{align}
T^{ab}_{ij} = W_{ij} \frac{(v_a-\varepsilon_j)(v_b-\varepsilon_i) + (v_a - \varepsilon_i)(v_b-\varepsilon_j)}{(\varepsilon_i - \varepsilon_j)(v_b-v_a)} - 2V_{iijj} \frac{(v_b-\varepsilon_i)(v_a-\varepsilon_i)}{(\varepsilon_i-\varepsilon_j)(v_b-v_a)}.
\end{align}
We can differentiate directly with respect to $\{\varepsilon\},g$
\begin{align} \label{eq:dEdeps}
\frac{\partial E}{\partial \varepsilon_k} &= 2\sum_{ja} \left(h_{jj} + \frac{1}{2} V_{jjjj}\right)\frac{\partial^2 v_a}{\partial \varepsilon_k \partial \varepsilon_j}
+ \sum_{\substack{ i\neq j \\ a<b}} \frac{\partial T^{ab}_{ij}}{\partial \varepsilon_k} \left( \frac{\partial v_a}{\partial \varepsilon_i} \frac{\partial v_b}{\partial \varepsilon_j} - \frac{\partial v_a}{\partial \varepsilon_j} \frac{\partial v_b}{\partial \varepsilon_i} \right) \nonumber \\
&+ \sum_{i\neq j} V_{iijj} \sum_a \left(  \frac{(\varepsilon_i-\varepsilon_j)\frac{\partial v_a}{\partial \varepsilon_k} -(v_a-\varepsilon_j)\delta_{ik}+(v_a-\varepsilon_i)\delta_{jk} }{(v_a-\varepsilon_j)^2} \frac{\partial v_a}{\partial \varepsilon_i} 
+ \frac{v_a-\varepsilon_i}{v_a-\varepsilon_j} \frac{\partial^2 v_a}{\partial \varepsilon_k \partial \varepsilon_i} \right) \nonumber \\
&+ \sum_{\substack{ i\neq j \\ a<b}} T^{ab}_{ij} \left( 
   \frac{\partial^2 v_a}{\partial \varepsilon_k \partial \varepsilon_i} \frac{\partial v_b}{\partial \varepsilon_j} +\frac{\partial v_a}{\partial \varepsilon_i} \frac{\partial^2 v_b}{\partial \varepsilon_k \partial \varepsilon_j}
- \frac{\partial^2 v_a}{\partial \varepsilon_k \partial \varepsilon_j} \frac{\partial v_b}{\partial \varepsilon_i} - \frac{\partial v_a}{\partial \varepsilon_j} \frac{\partial^2 v_b}{\partial \varepsilon_k \partial \varepsilon_i}
\right)
\end{align}
and
\begin{align} \label{eq:dEdg}
\frac{\partial E}{\partial g} &= 2\sum_{ja} \left(h_{jj} + \frac{1}{2} V_{jjjj}\right)\frac{\partial^2 v_a}{\partial g \partial \varepsilon_j} 
+ \sum_{i\neq j} V_{iijj} \sum_a \left( \frac{\varepsilon_i - \varepsilon_j}{(v_a-\varepsilon_j)^2} \frac{\partial v_a}{\partial g} \frac{\partial v_a}{\partial \varepsilon_i} 
+ \frac{v_a - \varepsilon_i}{v_a - \varepsilon_j} \frac{\partial^2 v_a}{\partial g \partial \varepsilon_i} \right) \nonumber \\
&+ \sum_{\substack{ i\neq j \\ a<b}} \frac{\partial T^{ab}_{ij}}{\partial g} \left( \frac{\partial v_a}{\partial \varepsilon_i} \frac{\partial v_b}{\partial \varepsilon_j} - \frac{\partial v_a}{\partial \varepsilon_j} \frac{\partial v_b}{\partial \varepsilon_i} \right)
+ \sum_{\substack{ i\neq j \\ a<b}} T^{ab}_{ij} \left( 
   \frac{\partial^2 v_a}{\partial g \partial \varepsilon_i} \frac{\partial v_b}{\partial \varepsilon_j} + \frac{\partial v_a}{\partial \varepsilon_i} \frac{\partial^2 v_b}{\partial g \partial \varepsilon_j}
- \frac{\partial^2 v_a}{\partial g \partial \varepsilon_j} \frac{\partial v_b}{\partial \varepsilon_i} -  \frac{\partial v_a}{\partial \varepsilon_j} \frac{\partial^2 v_b}{\partial g \partial \varepsilon_i}
\right)
\end{align}
with
\begin{align}
\frac{\partial T^{ab}_{ij}}{\partial \varepsilon_k} &=
2W_{ij}
\frac{(v_b-\varepsilon_i)(v_b-\varepsilon_j) \frac{\partial v_a}{\partial \varepsilon_k} - (v_a-\varepsilon_i)(v_a-\varepsilon_j) \frac{\partial v_b}{\partial \varepsilon_k} }{(\varepsilon_i - \varepsilon_j)(v_b-v_a)^2}
-2V_{iijj} \frac{(v_b-\varepsilon_i)^2 \frac{\partial v_a}{\partial \varepsilon_k} - (v_a-\varepsilon_i)^2 \frac{\partial v_b}{\partial \varepsilon_k}}{(\varepsilon_i - \varepsilon_j)(v_b-v_a)^2} \nonumber \\
&+ 2W_{ij} \frac{(v_a-\varepsilon_i)(v_b-\varepsilon_i)\delta_{jk} - (v_a-\varepsilon_j)(v_b-\varepsilon_j)\delta_{ik}}{(\varepsilon_i-\varepsilon_j)^2(v_b-v_a)} \nonumber \\
&-2 V_{iijj} \left(
\frac{\delta_{ik}}{(v_b-v_a)} + \frac{(v_a-\varepsilon_j) \left( (v_b-\varepsilon_i)\delta_{jk} - (v_b-\varepsilon_j)\delta_{ik} \right)}{(\varepsilon_i - \varepsilon_j)^2(v_b-v_a)}
\right)
\end{align}
\begin{align}
\frac{\partial T^{ab}_{ij}}{\partial g} = 2 W_{ij} \frac{(v_b-\varepsilon_i)(v_b-\varepsilon_j) \frac{\partial v_a}{\partial g} -(v_a-\varepsilon_i)(v_a-\varepsilon_j)\frac{\partial v_b}{\partial g}}{(\varepsilon_i - \varepsilon_j)(v_b-v_a)^2}
- 2 V_{iijj} \frac{(v_b-\varepsilon_i)^2 \frac{\partial v_a}{\partial g} - (v_a-\varepsilon_i)^2\frac{\partial v_b}{\partial g}}{(\varepsilon_i - \varepsilon_j)(v_b-v_a)^2}
\end{align}
Thus, to evaluate the elements of the gradient, all that is required are the second derivatives of the rapidities with respect to $\{\varepsilon\}$, as well as first and mixed derivatives with respect to the pairing strength $g$. First derivatives with $g$ can be evaluated in the same manner as those with respect to $\{\varepsilon\}$. Taking the derivative of Richardson's equations with respect to $g$ gives, for each $a$,
\begin{align}
0 = \left( \sum_{b\neq a} \frac{2}{(v_b-v_a)^2} - \sum_i \frac{1}{(v_a-\varepsilon_i)^2} \right) \frac{\partial v_a}{\partial g} - \sum_{b\neq a} \frac{2}{(v_b-v_a)^2} \frac{\partial v_b}{\partial g} - \frac{2}{g^2}
\end{align}
which can be seen as a set of linear equations
\begin{align} \label{eq:dudg}
G \frac{\partial \textbf{v}}{\partial g} = \textbf{p}
\end{align}
with the elements of the vector $\textbf{p}$ are the constant:
\begin{align}
p_a = - \frac{2}{g^2}
\end{align}
The matrix $G$ in \eqref{eq:dudg} is again the Gaudin matrix \eqref{eq:gmat}. The other derivatives required are evaluated in exactly the same manner. Second derivatives of Richardson's equations are listed in the appendix. Specifically, we solve the sets of linear equations:
\begin{align}
G \frac{\partial^2 \textbf{v}}{\partial \varepsilon_k^2} = \textbf{q}_k \\
G \frac{\partial^2 \textbf{v}}{\partial \varepsilon_l \partial \varepsilon_k} = \textbf{r}_{kl} \\
G \frac{\partial^2 \textbf{v}}{\partial g \partial \varepsilon_k} = \textbf{s}_k
\end{align}
with the $a$th elements of each of the RHSs:
\begin{align}
q_{k,a} &= 2 \left( \left( \sum_i \frac{1}{(v_a-\varepsilon_i)^3} + \sum_{b\neq a} \frac{2}{(v_b-v_a)^3} \right) \frac{\partial v_a}{\partial \varepsilon_k} - \sum_{b\neq a} \frac{2}{(v_b-v_a)^3} \frac{\partial v_b}{\partial \varepsilon_k} - \frac{1}{(v_a-\varepsilon_k)^3}
\right) \frac{\partial v_a}{\partial \varepsilon_k} \nonumber \\
&+ \sum_{b \neq a} \frac{4}{(v_b-v_a)^3}\frac{\partial v_b}{\partial \varepsilon_k} \frac{\partial v_b}{\partial \varepsilon_k}
- \sum_{b \neq a} \frac{4}{(v_b-v_a)^3} \frac{\partial v_a}{\partial \varepsilon_k} \frac{\partial v_b}{\partial \varepsilon_k}
- \frac{2}{(v_a - \varepsilon_k)^3} \frac{\partial v_a}{\partial \varepsilon_k} + \frac{2}{(v_a-\varepsilon_k)^3}
\end{align}
\begin{align}
r_{kl,a} &= 2 \left( \left( \sum_i \frac{1}{(v_a-\varepsilon_i)^3} + \sum_{b\neq a} \frac{2}{(v_b-v_a)^3} \right) \frac{\partial v_a}{\partial \varepsilon_l} - \sum_{b\neq a} \frac{2}{(v_b-v_a)^3} \frac{\partial v_b}{\partial \varepsilon_l} - \frac{1}{(v_a-\varepsilon_l)^3}
\right) \frac{\partial v_a}{\partial \varepsilon_k} \nonumber \\
&+ \sum_{b \neq a} \frac{4}{(v_b-v_a)^3}\frac{\partial v_b}{\partial \varepsilon_l} \frac{\partial v_b}{\partial \varepsilon_k}
- \sum_{b \neq a} \frac{4}{(v_b-v_a)^3} \frac{\partial v_a}{\partial \varepsilon_l} \frac{\partial v_b}{\partial \varepsilon_k}
- \frac{2}{(v_a - \varepsilon_k)^3} \frac{\partial v_a}{\partial \varepsilon_l}
\end{align}
\begin{align}
s_{k,a} &= 2 \left( \left( \sum_i \frac{1}{(v_a-\varepsilon_i)^3} + \sum_{b\neq a} \frac{2}{(v_b-v_a)^3} \right) \frac{\partial v_a}{\partial g} - \sum_{b\neq a} \frac{2}{(v_b-v_a)^3} \frac{\partial v_b}{\partial g}
\right) \frac{\partial v_a}{\partial \varepsilon_k} \nonumber \\
&+ \sum_{b \neq a} \frac{4}{(v_b-v_a)^3}\frac{\partial v_b}{\partial g} \frac{\partial v_b}{\partial \varepsilon_k}
- \sum_{b \neq a} \frac{4}{(v_b-v_a)^3} \frac{\partial v_a}{\partial g} \frac{\partial v_b}{\partial \varepsilon_k}
- \frac{2}{(v_a - \varepsilon_k)^3} \frac{\partial v_a}{\partial g}
\end{align}

The sets of linear equations can be solved numerically for all the required derivatives and the gradient can then be easily constructed. The gradient may be computed with a scaling of $\mathcal{O}(N^2M^3)$. The energy expression \eqref{eq:AO_energy} is much more efficient than that previously reported, and use of the analytic gradient allows us to employ conjugate gradient and quasi-Newton methods. Previously, we employed the Nelder-Mead simplex algorithm\cite{neldermead} to minimize the energy, which technically has better scaling than gradient based methods as it requires only evaluating the energy. However, Nelder-Mead requires orders of magnitude more iterations to converge than either conjugate gradient or quasi-newton approaches. It is of course understood that a good initial guess for the variational parameters is required.

\section{Reduced Density Matrix Elements: Gaudin Basis}
\label{sec:mobasis}
Rather than work with the local physical operators, we can work in the basis of Bethe ansatz quasiparticles: the pairs defined by the solutions of Richardson's equations. In quantum chemistry, this is analogous to working in the canonical Hartree-Fock molecular orbitals. The RG pairs are a linear transformation of the original local pair operators, though the transformation is rectangular:
\begin{align}
\begin{pmatrix}
S^+(v_1) \\
S^+(v_2) \\
\vdots \\
S^+(v_M)
\end{pmatrix} =
\begin{pmatrix}
\frac{1}{v_1 - \varepsilon_1} & \frac{1}{v_1 - \varepsilon_2} & \dots & \frac{1}{v_1 - \varepsilon_N} \\
\frac{1}{v_2 - \varepsilon_1} & \frac{1}{v_2 - \varepsilon_2} & \dots & \frac{1}{v_2 - \varepsilon_N} \\
\vdots & \vdots & \ddots & \vdots \\
\frac{1}{v_M - \varepsilon_1} & \frac{1}{v_M - \varepsilon_2} & \dots & \frac{1}{v_M - \varepsilon_N} \\
\end{pmatrix}
\begin{pmatrix}
S^+_1 \\
S^+_2 \\
\vdots \\
S^+_N
\end{pmatrix}
\end{align}
or
\begin{align} \label{eq:rect}
S^+(\mathbf{v}) = C \mathbf{S}^+.
\end{align}
The matrix $C$ has more columns than rows, and its rows are linearly independent. Therefore, $C$ has a right-inverse $C^R$, such that $CC^R=I_M$. Further, the explicit structure of $C^R$ is known,\cite{Schechter:1959} with elements
\begin{align}
[C^R]_{ia} = (v_a - \varepsilon_i) \prod_{k \neq i} \left( \frac{v_a - \varepsilon_k}{\varepsilon_i - \varepsilon_k} \right) \prod_{b \neq a} \left( \frac{\varepsilon_i - v_b}{v_a - v_b} \right).
\end{align}
The system \eqref{eq:rect} is under-determined, and thus has general solution:
\begin{align}
\mathbf{S}^+ = C^R S^+(\mathbf{v}) + \left( I_N - C^RC \right) \mathbf{w}_+
\end{align}
with $\mathbf{w}_+$ an arbitrary vector. The important point is that \emph{a} solution exists, and is unique up to the vector $\mathbf{w}_+$, for which we will make the choice $\mathbf{w}_+=0$ to retain clean expressions. Similarly,
\begin{align}
\mathbf{S}^- = C^R S^-(\mathbf{v}) + \left( I_N - C^RC \right) \mathbf{w}_-
\end{align}
The final element of the Gaudin algebra \eqref{eq:galg}, $S^z(u)$, can be written
\begin{align}
S^z(u) = \alpha(u) - \frac{1}{2} \sum_i \frac{\hat{n}_i}{u - \varepsilon_i},
\end{align}
so that 
\begin{align}
\hat{\mathbf{n}} = -2 C^R \left( S^z(\mathbf{u}) - \alpha(\mathbf{u}) \right) + \left( I_N - C^RC \right) \mathbf{w}_z.
\end{align}
The Hamiltonian \eqref{eq:sen0_ham} can be transformed to the Gaudin algebra:
\begin{align}
\hat{H}_0 = E_0 + \sum_a \tilde{h}_{aa} S^z(u_a) + \sum_{ab} \tilde{W}^z_{ab} S^z(u_a) S^z(u_b) + \sum_{ab} \tilde{W}^p_{ab} S^+(u_a) S^-(u_b)
\end{align}
with
\begin{align}
E_0 &= 2 \sum_{ia} h_{ii} C^R_{ia} \alpha(u_a) + \sum_{ijab} W_{ij} C^R_{ia} C^R_{jb} \alpha(u_a) \alpha(u_b) \\
\tilde{h}_{aa} &= -2  \sum_i h_{ii} C^R_{ia} -2 \sum_{ijb}W_{ij}C^R_{ia}C^R_{jb} \alpha(u_b) \\
\tilde{W}^z_{ab} &= \sum_{ij} W_{ij} C^R_{ia} C^R_{jb} \\
\tilde{W}^p_{ab} &= \sum_{ij} V_{iijj} C^R_{ia} C^R_{jb}.
\end{align}
Transforming the Hamiltonian to the GB requires only double sums over each set of elements, and thus may be performed naively with $\mathcal{O}(N^2M^2)$ scaling, though by performing the summations sequentially and saving the intermediates this is easily reduced.

The energy expression becomes
\begin{align}
E = E_0 + \sum_a h_a \tilde{Z}_a + \sum_{ab} \tilde{W}^z_{ab} \tilde{Z}_{ab} + \tilde{W}^p_{ab} \tilde{P}_{ab}
\end{align}
with the RDM elements (or correlation functions) defined:
\begin{align}
\tilde{Z}_a &=     \frac{ \braket{\{v\} | S^z(v_a)| \{v\}}} {\braket{\{v\}|\{v\}}} \\
\tilde{Z}_{ab} &= \frac{ \braket{\{v\} | S^z(v_a) S^z(v_b) | \{v\}}} {\braket{\{v\}|\{v\}}} \\
\tilde{P}_{ab} &= \frac{ \braket{\{v\} | S^+(v_a) S^-(v_b) | \{v\}}} {\braket{\{v\}|\{v\}}}.
\end{align}
These RDM elements have simple expressions in terms of sums of determinants that may be evaluated efficiently with Cramer's rule. With the results of the previous section, along with the definitions of the Gaudin algebra, it is in principle possible to employ determinant identities to show this directly. We will instead use the Gaudin algebra, acting on RG eigenvectors, to arrive at this result in a more illuminating manner. The development parallels the approach in the physical basis: we will act with the algebraic objects on the eigenvectors to reduce the RDMs to sums of specific scalar products that may be evaluated as limiting cases of Slavnov's theorem. As working in the Gaudin algebra basis is more difficult than the physical basis, and to our knowledge hasn't been done (though Sklyanin has thought through these lines\cite{sklyanin:1999}), we report more intermediate stages. Actions of the Gaudin algebra on RG eigenvectors will be useful in following papers to construct transition density matrices.

\subsection{Gaudin algebra actions}
For an arbitrary $u$, and $\{v\}$ a solution of Richardson's equations, we will use the definition:
\begin{align}
f(u) = \alpha(u) - \sum_{b} \frac{1}{u - v_b}
\end{align}
to write the action of $S^z(u)$ on an eigenvector:
\begin{align}
S^z(u) \ket{ \{v\} } = \sum_a \frac{S^+(u)}{u-v_a} \ket{ \{v\}_a } + f(u) \ket{ \{v\} }
\end{align}
We can take the limit $u\rightarrow v_a$ for one the rapidities, which after using L'hopital's rule, yields:
\begin{align} \label{eq:Za}
S^z (v_a) \ket{ \{v\}} = \frac{\partial S^+ (v_a)}{\partial v_a} \ket{ \{v\}_a} + \sum_{b \neq a} \frac{S^+ (v_a)}{v_a -v_b} \ket{ \{v\}_b}
\end{align}

For arbitrary $u_1,u_2$, the action of $S^z(u_1)S^z(u_2)$ yields
\begin{align}
S^z(u_1) S^z(u_2)\ket{\{v\}} &= \sum_{a\neq b} \left( \frac{1}{(u_1 - v_a)(u_2 - v_b)} + \frac{1}{(u_1-v_b)(u_2-v_a)} \right) S^+(u_1) S^+(u_2) \ket{ \{v\}_{a,b} } \nonumber \\
&+ \sum_a \left( \frac{f(u_2)}{u_1-v_a}  - \frac{1}{(u_2-v_a)(u_2-u_1)} \right) S^+ (u_1) \ket{\{v\}_a} \nonumber \\
&+ \sum_a \left( \frac{f(u_1)}{u_2-v_a}  - \frac{1}{(u_1-v_a)(u_1-u_2)} \right) S^+ (u_2) \ket{\{v\}_a} \nonumber \\
&+ f(u_1) f(u_2) \ket{ \{v\}}
\end{align}
In the limit of solutions of Richardson's equations, for $v_a \neq v_b$, the action becomes
\begin{align} \label{eq:Zab}
S^z(v_a) S^z(v_b) \ket{\{v\}} &= \frac{\partial S^+(v_a)}{\partial v_a} \frac{\partial S^+(v_b)}{\partial v_b} \ket{\{v\}_{a,b}}  - \frac{3}{(v_a-v_b)^2} \ket{\{v\}} \nonumber \\
&+ \sum_{c\neq a,b} \left( \frac{1}{v_b-v_c}\frac{\partial S^+(v_a)}{\partial v_a}S^+(v_b) \ket{\{v\}_{a,c}} + \frac{1}{v_a-v_c}\frac{\partial S^+(v_b)}{\partial v_b}S^+(v_a) \ket{\{v\}_{b,c}} \right) \nonumber \\
&+ \sum_{c\neq a,b} \left( \frac{2}{(v_a-v_b)(v_b-v_c)}S^+(v_a)\ket{\{v\}_c} + \frac{2}{(v_b-v_a)(v_a-v_c)} S^+(v_b) \ket{\{v\}_c} \right) \nonumber \\
&+ \frac{1}{2} \sum_{c,d \neq a,b} \left( \frac{1}{(v_a - v_c)(v_d-v_b)} + \frac{1}{(v_a-v_d)(v_b-v_c)} \right) S^+(v_a)S^+(v_b) \ket{\{v\}_{c,d}} \nonumber \\
&+ \frac{1}{(v_a-v_b)^2} S^+(v_a) \ket{\{v\}_b} + \frac{1}{(v_a-v_b)^2} S^+(v_b) \ket{\{v\}_a}.
\end{align}
For the diagonal element, the result is
\begin{align} \label{eq:Zaa}
S^z(v_a)S^z(v_a) \ket{\{v\}} &= \frac{1}{2} \frac{\partial^2 S^+(v_a)}{\partial v_a^2} \ket{\{v\}_a} + \sum_{c,d \neq a} \frac{S^+(v_a)S^+(v_a)}{(v_a-v_c)(v_a-v_d)} \ket{\{v\}_{c,d}} \nonumber \\
&+ \sum_{c\neq a} \frac{3}{v_a-v_c} \frac{\partial S^+(v_a)}{\partial v_a} \ket{\{v\}_c} + \frac{S^+(v_a)}{(v_a-v_c)^2} \ket{\{v\}_c}
\end{align}

Likewise, the $S^+(u_1)S^-(u_2)$ action is simplified with the shorthand:
\begin{align}
f_a (u) = \alpha (u) - \sum_{b\neq a} \frac{1}{u - v_b}
\end{align}
\begin{align}
S^+(u_1) S^-(u_2)\ket{\{v\}}  
&= -2\sum_a \frac{f_a (u_2) - f_a (v_a)}{u_2-v_a}  S^+(u_1) \ket{\{v\}_a} -2 \sum_{b \neq a} \frac{S^+(u_1) S^+(u_2)}{(v_a - u_2)(v_b - u_2)}  \ket{ \{v\}_{a,b}}
\end{align}
Which becomes, for $a \neq b$:
\begin{align} \label{eq:Pab}
S^+(v_a) S^-(v_b) \ket{\{v\}} &= -2 \left( \frac{\partial \alpha (v_b)}{\partial v_b } +  \sum_{c\neq b} \frac{1}{(v_c - v_b)^2} \right) S^+(v_a) \ket{\{v\}_b} \nonumber \\
&+ 2 \sum_{c \neq b} \frac{S^+(v_a)}{v_c - v_b}  \frac{\partial S^+(v_b)}{\partial v_b} \ket{\{v\}_{b,c}} 
- 2 \sum_{c\neq b} \frac{S^+(v_a)}{(v_c - v_b)^2}  \ket{\{v\}_c} \nonumber \\
&- \sum_{c,d\neq b} \frac{S^+(v_a)S^+(v_b)}{(v_b-v_c)(v_b-v_d)}  \ket{ \{v\}_{c,d} }
\end{align}
and for $a=b$
\begin{align} \label{eq:Paa}
S^+(v_a) S^-(v_a) \ket{\{v\}} &= -2 \left( \frac{\partial \alpha (v_a)}{\partial v_a } +  \sum_{c\neq a} \frac{1}{(v_c - v_a)^2} \right)  \ket{\{v\}} \nonumber \\
&+ 2 \sum_{c \neq a} \frac{1}{v_c - v_a}  \frac{\partial S^+(v_a)}{\partial v_a} \ket{\{v\}_{c}} 
- 2 \sum_{c\neq a} \frac{S^+(v_a)}{(v_c - v_a)^2}  \ket{\{v\}_c} \nonumber \\
&- \sum_{c,d\neq a} \frac{S^+(v_a)S^+(v_a)}{(v_a-v_c)(v_a-v_d)}  \ket{ \{v\}_{c,d} } .
\end{align}

For consistency, we can verify these results by looking at the action of the transfer matrix $S^2(u)$ upon an eigenvector in the limit that $u$ becomes one of the rapidities. For this purpose, we will require the action of the derivative of $S^z(u)$ upon an eigenvector
\begin{align}
\frac{\partial S^z(u)}{\partial u} \ket{\{v\}}= \sum_a \frac{1}{u-v_a} \left( \frac{\partial S^+(u)}{\partial u} - \frac{S^+(u)}{u-v_a}  \right) \ket{\{v\}_a} + \frac{\partial f(u)}{\partial u} \ket{\{v\}}
\end{align}
which, for one of the rapidities yields
\begin{align}
\frac{\partial S^z(v_a)}{\partial v_a} \ket{\{v\}}&=  \frac{\partial f_a(v_a)}{\partial v_a} \ket{\{v\}} + \frac{1}{2} \frac{\partial^2 S^+(v_a)}{\partial v_a^2} \ket{ \{v\}_a} \nonumber \\
&+ \sum_{b \neq a} \frac{1}{v_a-v_b}\left( \frac{\partial S^+(v_a)}{\partial v_a} - \frac{S^+(v_a)}{(v_a-v_b)} \right) \ket{\{v\}_b}.
\end{align}
With these results, we can verify the action of the transfer matrix eigenvalue evaluated at one of the rapidities:
\begin{align} \label{eq:Da}
S^2(v_a) \ket{\{v\}} &= \left(S^z(v_a) S^z(v_a) +S^+(v_a)S^-(v_a) - \frac{\partial S^z(v_a)}{\partial v_a} \right)\ket{\{v\}} \\
&= -3 \left( \frac{\partial \alpha(v_a)}{\partial v_a} + \sum_{b \neq a} \frac{1}{(v_a-v_b)^2}      \right) \ket	{\{v\}} \label{eq:s2con}.
\end{align}
In equation \eqref{eq:s2con}, the factor multiplying $\ket{\{v\}}$ is precisely the eigenvalue $\Lambda \left( u, \{v\} \right)$ in the limit $u\rightarrow v_a$. In particular it is equal to $\frac{3}{2}G_{aa}$ where $G_{aa}$ are the diagonal elements of the Gaudin matrix.

\subsection{Form factors}
With the results of the previous section, we now evaluate form factors. In the local basis, there were only two possible form factors to evaluate, corresponding to residues of simple poles of Slavnov's determinant. The results were further simplified as solutions of $N$ sets of linear equations by Cramer's rule. In the GB there are eight form factors to evaluate. Remarkably, they may all be computed with the solutions of $2 M$ sets of linear equations.  Specifically, the form factors required are expressible in terms of the vectors $\textbf{t}^{(1)}_a$ and $\textbf{t}^{(2)}_a$, with $b$th elements:
\begin{align}
t^{(1)}_{a,b} = 
\begin{cases}
\sum_i  \frac{1}{(v_a - \varepsilon_i)^3} - \sum_{c \neq a} \frac{2}{(v_a - v_c)^3} \quad &b=a \\
\frac{6}{(v_a-v_b)^3} \quad &b\neq a
\end{cases}
\end{align}
\begin{align}
t^{(2)}_{a,b} = 
\begin{cases}
\sum_i \frac{1}{(v_a - \varepsilon_i)^4} - \sum_{c \neq a} \frac{2}{(v_a - v_c)^4} \quad &b=a \\
\frac{12}{(v_a - v_b)^4} + \frac{1}{(v_a -v_b)^2} \left( \sum_i \frac{1}{(v_a - \varepsilon_i)^2} - \sum_{c\neq a} \frac{2}{(v_a -v_c)^2} \right) \quad & b \neq a
\end{cases}
\end{align}
There are $M$ different $\textbf{t}^{(1)}_a$ and $M$ different $\textbf{t}^{(2)}_a$, one each for each rapidity in the ground state of Richardson's equations. In the form factor expressions, the notation $\det G^{a}_c$ will represent the Gaudin matrix with the $c$th column replaced with the vector $\textbf{t}^{(1)}_a$ while $\det G^{\bar{a}}_c$ means the Gaudin matrix with the $c$th column replaced with the vector $\textbf{t}^{(2)}_a$. 

To evaluate form factors with repeated arguments, all that is required is to take appropriate limits of Slavnov's theorem. Generally, we have found the easiest manner to accomplish this is to begin with $\{v\}$ a solution of Richardson's equations and $\{u\}$ arbitrary. Next set $u_a = v_a + h$ for some small $h$ and expand the result with the geometric series in powers of $h$. Terms proportional to negative powers of $h$ all vanish identically, and those with positive powers will vanish in the limit $h\rightarrow 0$. The remaining, desired, terms are the result. In particular to evaluate $\braket{ \{v\} | S^+(v_a) | \{v\}_c}$, start with Slavnov's theorem for $\braket{ \{v\} | \{u\}}$ and take the limit $\{u\} \rightarrow \{v\}$ for all $\{u\}$ except $u_c$. Then, set $u_c = v_a +h$ and expand each of the rational terms as a geometric series, eg.
\begin{align}
\frac{1}{(v_a +h - \varepsilon_i)} = \frac{1}{(v_a - \varepsilon_i)} \frac{1}{\left( 1 - \frac{-h}{(v_a-\varepsilon_i)} \right)}= \frac{1}{(v_a - \varepsilon_i)} \left( 1 - \frac{h}{(v_a - \varepsilon_i)} + \frac{h^2}{(v_a-\varepsilon_i)^2} + \dots \right)
\end{align}
and collect terms. The one term proportional to $h^{-1}$ is the determinant of $G$ with the $a$th column repeated, and is hence identically zero. Terms proportional to positive powers of $h$ vanish as $h\rightarrow 0$, and the remaining term is
\begin{align} \label{eq:ff1}
\braket{ \{v\} | S^+(v_a) | \{v\}_c} &= (v_c - v_a) \det G^{a}_c \\
&= \sum_i \frac{K^i_c}{v_a - \varepsilon_i}
\end{align}
where, in the second equality
\begin{align} \label{eq:PB1}
K^i_c = (v_c - \varepsilon_i) \det G^i_c 
\end{align}
is a direct transformation of the result in the PB we have used as a numerical consistency check. In what follows, each form factor will have two expressions, the first being the desired expression in the GB, and the second being a direct transformation of the PB results \eqref{eq:PB1} and \eqref{eq:PB2} to be used as a numerical check
\begin{align} \label{eq:PB2}
K^{ij}_{ab} = \frac{(v_a-\varepsilon_i)(v_a-\varepsilon_j)(v_b-\varepsilon_i)(v_b-\varepsilon_j)}{(\varepsilon_i - \varepsilon_j)(v_b-v_a)}\det G^{ij}_{ab}.
\end{align}
Each form factor result has been verified numerically for a variety of reduced BCS Hamiltonians.

Double replacement form factors are calculated in the same manner as \eqref{eq:ff1}, the results being
\begin{align} \label{eq:ff2}
\braket{ \{v\} | S^+(v_a) S^+(v_a) | \{v\}_{c,d}} &= \frac{(v_c -v_a)^2 (v_d - v_a)^2}{v_d - v_c} \det G^{a \bar{a}}_{cd} \\
&= \sum_{ij} \frac{K^{ij}_{cd}}{(v_a-\varepsilon_i)(v_a - \varepsilon_j)}
\end{align}
and
\begin{align} \label{eq:ff3}
\braket{ \{v\} | S^+(v_a) S^+(v_b) | \{v\}_{c,d}} &= \frac{(v_c -v_a)(v_d-v_a)(v_c-v_b)(v_d-v_b)}{(v_a-v_b)(v_d-v_c)} \det G^{ab}_{cd} \\
&= \sum_{ij} \frac{K^{ij}_{cd}}{(v_a-\varepsilon_i)(v_b-\varepsilon_j)}.
\end{align}

Form factors involving derivatives $\frac{\partial S^+(v)}{\partial v}$ may be evaluated by taking the appropriate derivative of Slavnov's theorem, then taking the limit $\{u\} \rightarrow \{v\}$. For example, begin by taking the derivative of Slavnov's theorem with respect to $u_a$
\begin{align} 
\braket{ \{v\} | \frac{\partial S^+(u_a)}{\partial u_a} | \{u\}_a} &= \frac{\partial }{\partial u_a } \braket{\{v\}|\{u\}} \nonumber \\ 
&= \frac{\partial K}{\partial u_a} \det J + K \frac{\partial \det J}{\partial u_a}.
\end{align}
The derivative of $K$ simplifies
\begin{align}
\frac{\partial K}{\partial u_a} = Kk(u_a),
\end{align}
with
\begin{align}
k(u_a) = \sum_b \frac{1}{u_a - v_b} - \sum_{c \neq a} \frac{1}{u_a - u_c}.
\end{align}
The derivative of the determinant simplifies as only the $a$th column depends on $u_a$:
\begin{align}
\frac{\partial \det J}{\partial u_a} = \det (J_1 | \dots |  \frac{\partial J_a}{\partial u_a}   | \dots| J_{M})
\end{align}
and the expression becomes:
\begin{align}
\braket{ \{v\} | \frac{\partial S^+(u_a)}{\partial u_a} | \{u\}_a} = K \det (J_1 | \dots | k(u_a)  J_a +  \frac{\partial J_a}{\partial u_a}  | \dots| J_{M}).
\end{align}
Taking the limit $\{u\} \rightarrow \{v\}$ gives:
\begin{align} \label{eq:ff4}
\braket{ \{v\} | \frac{\partial S^+(v_a)}{\partial v_a} | \{v\}_a} &= - \det G^{a}_a \\
&= -\sum_i \frac{K^i_a}{(v_a-\varepsilon_i)^2}.
\end{align}
The same approach gives
\begin{align} \label{eq:ff5}
\braket{ \{v\} | \frac{\partial S^+(v_a)}{\partial v_a} | \{v\}_c} &= -\det G^{a}_c - (v_c - v_a) \det G^{\bar{a}}_c \\
&= - \sum_i \frac{K^i_c}{(v_a-\varepsilon_i)^2},
\end{align}

\begin{align} \label{eq:ff6}
\braket{ \{v\} | \frac{\partial S^+(v_a)}{\partial v_a} S^+(v_b)| \{v\}_{a,c}} &= \frac{(v_c-v_b)^2}{(v_a-v_b)(v_a-v_c)}\det G^{b}_c - (v_c -v_b)\det G^{ab}_{ac} \\
&= -\sum_{ij} \frac{K^{ij}_{ac}}{(v_a-\varepsilon_i)^2(v_b-\varepsilon_j)}
\end{align}
and
\begin{align} \label{eq:ff7}
\braket{ \{v\} | \frac{\partial S^+(v_a)}{\partial v_a} \frac{\partial S^+(v_b)}{\partial v_b} | \{v\}_{a,b}} &= \det G^{ab}_{ab} - \frac{1}{(v_a -v_b)^2} \det G \\
&= \sum_{ij} \frac{K^{ij}_{ab}}{(v_a-\varepsilon_i)^2(v_b-\varepsilon_j)^2}.
\end{align}

The second derivative form factor is evaluated similarly. Starting with Slavnov's theorem, and taking a second derivative yields the expression
\begin{align}
\braket{ \{v\} | \frac{\partial^2 S^+(u_a)}{\partial u_a^2}  | \{u\}_a} = K \det (J_1 | \dots | \left(k(u_a)^2 + \frac{\partial k(u_a)}{\partial u_a}\right) J_a + 2k(u_a) \frac{\partial J_a}{\partial u_a} + \frac{\partial^2 J_a}{\partial u_a^2}  | \dots| J_{M})
\end{align}
where only the $a$th column is changed. Setting $\{u\}\rightarrow\{v\}$ yields
\begin{align} \label{eq:ff8}
\braket{ \{v\} | \frac{\partial^2 S^+(v_a)}{\partial v_a^2}  | \{v\}_a} &= 2 \det G^{\bar{a}}_a \\
&= 2\sum_i \frac{K^i_a}{(v_a-\varepsilon_i)^3}.
\end{align}

With all the form factor expressions in hand, we now move to the RDM elements.
\subsection{RDM elements}
In computing the normalized matrix elements, we will always have ratios of determinants, with the denominator the Gaudin matrix $G$. In this way, we can replace the ratios with the symbols:
\begin{align}
X^{a}_c &= \frac{\det G^{a}_c}{\det G} \\
Y^{a}_c &= \frac{\det G^{\bar{a}}_c}{\det G}
\end{align}
Again, from Cramer's rule, these objects are computed as solutions of the $2M$ sets of linear equations:
\begin{align}
G \mathbf{X}^a = \mathbf{t}^{(1)}_a \\
G \mathbf{Y}^a = \mathbf{t}^{(2)}_a
\end{align}

From \eqref{eq:Za}, \eqref{eq:ff1} and \eqref{eq:ff4}, $\tilde{Z}_a$ can be computed
\begin{align}
\tilde{Z}_a = - \sum_b X^a_b.
\end{align}
The simplicity of this result is again suggestive that it is optimal. The expressions for $\tilde{Z}_{ab}$ are less clean, but are easily computed, for the diagonal from Jacobi's theorem along with \eqref{eq:Zaa}, \eqref{eq:ff1}, \eqref{eq:ff2}, \eqref{eq:ff5} and \eqref{eq:ff8}
\begin{align}
\tilde{Z}_{aa} = Y^a_a + \sum_{c \neq a} \left( \frac{4X^a_c}{v_c-v_a} + 3Y^a_c \right)+ \sum_{c,d \neq a} \frac{(v_c-v_a)(v_d-v_a)}{v_d-v_c} \left( X^a_cY^a_d - X^a_dY^a_c \right)
\end{align}
while the off-diagonal is computed from \eqref{eq:Zab}, \eqref{eq:ff1}, \eqref{eq:ff3}, \eqref{eq:ff6} and \eqref{eq:ff7}
\begin{align}
\tilde{Z}_{ab} =& X^a_aX^b_b - X^a_bX^b_a + \frac{X^a_b}{v_b-v_a} + \frac{X^b_a}{v_a-v_b} -\frac{4}{(v_a-v_b)^2} \nonumber \\
&+ \sum_{c\neq a,b} \left( X^a_a X^b_c - X^a_cX^b_a + X^b_bX^a_c - X^b_cX^a_b \right) + \frac{3}{v_b-v_a} \left( \frac{(v_a-v_c)}{(v_b-v_c)} X^a_c - \frac{(v_b-v_c)}{(v_a-v_c)}X^b_c \right) \nonumber \\
&+ \frac{1}{2} \sum_{c,d \neq a,b} \frac{(v_d-v_a)(v_c-v_b) + (v_c-v_a)(v_d-v_b)}{(v_a-v_b)(v_d-v_c)} \left( X^a_cX^b_d - X^a_dX^b_c \right) .
\end{align}
The diagonal of $\tilde{P}_{ab}$ is computed from \eqref{eq:Paa}, \eqref{eq:ff1}, \eqref{eq:ff2} and \eqref{eq:ff5}:
\begin{align}
\tilde{P}_{aa} &= G_{aa} - 2 \sum_{c\neq a} \left(\frac{2X^a_c}{v_c - v_a} + Y^a_c \right) - \sum_{c,d\neq a} \frac{(v_c-v_a)(v_d-v_a)}{v_d-v_c} (X^a_cY^a_d - X^a_dY^a_c)  
\end{align}
while the off-diagonal is computed from \eqref{eq:Pab}, \eqref{eq:ff1}, \eqref{eq:ff3} and \eqref{eq:ff6}
\begin{align}
\tilde{P}_{ab} &= G_{bb} (v_b-v_a) X^a_b - \frac{2}{(v_a-v_b)^2}  - \sum_{c\neq a}\frac{2 X^b_c}{v_a-v_b} \nonumber \\
&-2 \sum_{c\neq a,b} \frac{v_c-v_a}{v_c-v_b} \left( X^b_bX^a_c - X^b_cX^a_b + \frac{X^a_c}{v_c-v_b} \left( 1 + \frac{v_c-v_a}{v_b-v_a} \right) \right) \nonumber \\
&+ \sum_{c,d \neq a,b} \frac{(v_c-v_a)(v_d-v_a)}{(v_b-v_a)(v_d-v_c)} \left( X^a_cX^b_d - X^a_dX^b_c \right).
\end{align}

\subsection{Consistency checks}
As the development of the RDM elements in the GB is rather long and prone to subtle errors, we have verified our formulas numerically by transforming the correlation functions from the PB directly. As mentioned earlier, we have also verified intermediate results for the form factors. The simplest RDM elements to verify are $\tilde{P}_{ab}$, as they are just
\begin{align}
\tilde{P}_{ab} = \sum_{ij} \frac{P_{ij}}{(v_a-\varepsilon_i)(v_b-\varepsilon_j)}
\end{align}
For $\tilde{Z}_{a}$, 
\begin{align}
\tilde{Z}_a &= \frac{1}{g} + \frac{1}{2} \sum_i \frac{1}{v_a -\varepsilon_i} - \sum_i \frac{\gamma_i}{v_a-\varepsilon_i} \\
&= \sum_{b \neq a} \frac{1}{v_a - v_b} - \sum_i \frac{\gamma_i}{v_a-\varepsilon_i}
\end{align}
where Richardson's equations have been used in the last line. Similarly for $\tilde{Z}_{ab}$,
\begin{align}
\tilde{Z}_{ab} &= \sum_{\substack{ c \neq a \\ d \neq b}} \frac{1}{(v_c - v_a)(v_d - v_b)} - \frac{1}{g} \sum_i \gamma_i \left( \frac{1}{v_a - \varepsilon_i} + \frac{1}{v_b - \varepsilon_i} \right)
+ \sum_{ij} \frac{D_{ij} - \frac{1}{2} \gamma_i - \frac{1}{2} \gamma_j}{(v_a-\varepsilon_i)(v_b-\varepsilon_j)}
\end{align}
Again, these results have been verified numerically for a variety of Hamiltonians \eqref{eq:BCSham} comprising weak and strong coupling limits.

We can also write sum rules in terms of the EBV \eqref{eq:ebv}
\begin{align}
\sum_a \tilde{Z}_{a} = \sum_i U_i \gamma_i
\end{align}
\begin{align}
\sum_{ab} \tilde{Z}_{ab} = \frac{2M}{g} \sum_i U_i \gamma_i + \sum_{ij} U_i U_j \left( D_{ij} - \frac{1}{2}\gamma_i - \frac{1}{2} \gamma_j \right)
\end{align}
\begin{align}
\sum_{ab} \tilde{P}_{ab} = \sum_{ij} U_i U_j P_{ij}
\end{align}
In the PB, the sum rules were all expressible in terms of physical quantities, whereas in the GB the EBV enter the formulae. 

With equations \eqref{eq:Da}, \eqref{eq:ff1}, \eqref{eq:ff5} and \eqref{eq:ff8} we can define
\begin{align}
\tilde{D}_a = \frac{\braket{ \{v\}| \frac{\partial S^z(v_a)}{\partial v_a} | \{v\}}}{\braket{\{v\}|\{v\}}} = - \frac{1}{2}G_{aa} + \sum_c Y^a_c
\end{align}
and verify the expectation value of $S^2(v_a)$:
\begin{align}
\frac{\braket{ \{v\}| S^2(v_a) | \{v\} }}{\braket{ \{v\} | \{v\} }} = \tilde{Z}_{aa} + \tilde{P}_{aa} - \tilde{D}_a = \frac{3}{2} G_{aa}
\end{align}
which is consistent with \eqref{eq:s2con}.

Finally, to verify our formulas, we have also used the partial summations for $\tilde{P}_{ab}$
\begin{align}
\sum_a \tilde{P}_{aa} = - \sum_{i\neq j} \frac{U_i - U_j}{\varepsilon_i - \varepsilon_j} + \sum_a \sum_i \frac{P_{ii}}{(v_a -\varepsilon_i)^2}
\end{align}
\begin{align}
\sum_b \tilde{P}_{ab} = \sum_{ij}  \frac{U_j P_{ij}}{\varepsilon_i - v_a}
\end{align}
and for $\tilde{Z}_{ab}$
\begin{align}
\sum_a \tilde{Z}_{aa} =  \sum_{a<b} \frac{2}{(v_a - v_b)^2} + \frac{2}{g} \sum_i \gamma_i U_i - \sum_{i \neq j} \left( D_{ij} - \frac{1}{2} \gamma_i - \frac{1}{2} \gamma_j \right) \frac{U_i - U_j}{\varepsilon_i - \varepsilon_j} 
\end{align}
\begin{align}
\sum_b \tilde{Z}_{ab} = \frac{1}{g} \sum_j U_j \gamma_j + \frac{1}{g} \sum_b \sum_i \frac{\gamma_i}{\varepsilon_i - v_a} + \sum_{ij}  \frac{U_j (D_{ij} - \frac{1}{2}\gamma_i - \frac{1}{2}\gamma_j)}{\varepsilon_i - v_a}.
\end{align}

\section{Summary of Principal Results} \label{sec:summary}
In this section, we summarize the information required to calculate RDMs in both the PB and the GB. With $\{v\}$ a solution of Richardson's equations and $\{\varepsilon \}$ the single particle energies defining a reduced BCS Hamiltonian, the Gaudin matrix is
\begin{align}
G_{ab} &= 
\begin{cases}
\sum_i \frac{1}{(v_a -\varepsilon_i)^2} - \sum_{c\neq a} \frac{2}{(v_a -v_a)^2}, \quad &a=b \\
\frac{2}{(v_a-v_b)^2}, \quad &a\neq b.
\end{cases}
\end{align}
This matrix will naturally become sparse as the off-diagonal elements go to zero rapidly. One can solve the sets of linear equations:
\begin{align}
G \frac{\partial \textbf{v}}{\partial \varepsilon_k} &= \textbf{b}_k \\
G \textbf{X}^a &= \textbf{t}^{(1)}_a \\
G \textbf{Y}^a &= \textbf{t}^{(2)}_a
\end{align}
with the $b$th element of the RHSs
\begin{align}
b_{k,b} &= \frac{1}{(v_b - \varepsilon_k)^2}\\
t^{(1)}_{a,b} &= \begin{cases}
\sum_i  \frac{1}{(v_a - \varepsilon_i)^3} - \sum_{c \neq a} \frac{2}{(v_a - v_c)^3} \quad &b=a \\
\frac{6}{(v_a-v_b)^3} \quad &b\neq a
\end{cases} \\
t^{(2)}_{a,b} &= \begin{cases}
\sum_i \frac{1}{(v_a - \varepsilon_i)^4} - \sum_{c \neq a} \frac{2}{(v_a - v_c)^4} \quad &b=a \\
\frac{12}{(v_a - v_b)^4} + \frac{1}{(v_a -v_b)^2} \left( \sum_i \frac{1}{(v_a - \varepsilon_i)^2} - \sum_{c\neq a} \frac{2}{(v_a -v_c)^2} \right) \quad & b \neq a.
\end{cases}
\end{align}
The RDMs in the PB are simple sums of the results:
\begin{align}
 \frac{1}{2} \langle \hat{n}_i \rangle  &= \sum_a \frac{\partial v_a}{\partial \varepsilon_i} \\
 \frac{1}{4} \langle \hat{n}_i \hat{n}_j \rangle &=\sum_{a<b} \frac{(v_a - \varepsilon_i)(v_b - \varepsilon_j) + (v_a - \varepsilon_j)(v_b - \varepsilon_i)}{(\varepsilon_i - \varepsilon_j)(v_b-v_a)} 
\left( \frac{\partial v_a}{\partial \varepsilon_i} \frac{\partial v_b}{\partial \varepsilon_j} - \frac{\partial v_a}{\partial \varepsilon_j} \frac{\partial v_b}{\partial \varepsilon_i} \right)  \\
 \frac{1}{2} \langle S^+_i S^-_j \rangle&= \sum_a \frac{v_a-\varepsilon_i}{v_a - \varepsilon_j} \frac{\partial v_a}{\partial \varepsilon_i} -2 \sum_{a<b} \frac{(v_b-\varepsilon_i)(v_a - \varepsilon_i)}{(\varepsilon_i - \varepsilon_j)(v_b-v_a)}
\left( \frac{\partial v_a}{\partial \varepsilon_i} \frac{\partial v_b}{\partial \varepsilon_j} - \frac{\partial v_a}{\partial \varepsilon_j} \frac{\partial v_b}{\partial \varepsilon_i} \right)
\end{align}
Thus after solving $N$ linear equations, each with cost $\mathcal{O}(M^3)$, the RDMs are easily computed. In the GB the results are:
\begin{align}
\langle S^z(v_a) \rangle &= - \sum_b X^a_b. \\
\langle  S^z(v_a) S^z(v_a) \rangle &= Y^a_a + \sum_{c \neq a} \left( \frac{4X^a_c}{v_c-v_a} + 3Y^a_c \right)+ \sum_{c,d \neq a} \frac{(v_c-v_a)(v_d-v_a)}{v_d-v_c} \left( X^a_cY^a_d - X^a_dY^a_c \right) \\
\langle  S^z(v_a) S^z(v_b) \rangle &= X^a_aX^b_b - X^a_bX^b_a + \frac{X^a_b}{v_b-v_a} + \frac{X^b_a}{v_a-v_b} -\frac{4}{(v_a-v_b)^2} \nonumber \\
&+ \sum_{c\neq a,b} \left( X^a_a X^b_c - X^a_cX^b_a + X^b_bX^a_c - X^b_cX^a_b \right) + \frac{3}{v_b-v_a} \left( \frac{v_a-v_c}{v_b-v_c} X^a_c - \frac{v_b-v_c}{v_a-v_c}X^b_c \right) \nonumber \\
&+ \frac{1}{2} \sum_{c,d \neq a,b} \frac{(v_d-v_a)(v_c-v_b) + (v_c-v_a)(v_d-v_b)}{(v_a-v_b)(v_d-v_c)} \left( X^a_cX^b_d - X^a_dX^b_c \right)\\
\langle  S^+(v_a) S^-(v_a) \rangle &= G_{aa} - 2 \sum_{c\neq a} \left(\frac{2X^a_c}{v_c - v_a} + Y^a_c \right) - \sum_{c,d\neq a} \frac{(v_c-v_a)(v_d-v_a)}{v_d-v_c} (X^a_cY^a_d - X^a_dY^a_c) \\
\langle  S^+(v_a) S^-(v_b) \rangle &= G_{bb} (v_b-v_a) X^a_b - \frac{2}{(v_a-v_b)^2}  - \sum_{c\neq a}\frac{2 X^b_c}{v_a-v_b} \nonumber \\
&-2 \sum_{c\neq a,b} \frac{v_c-v_a}{v_c-v_b} \left( X^b_bX^a_c - X^b_cX^a_b + \frac{X^a_c}{v_c-v_b} \left( 1 + \frac{v_c-v_a}{v_b-v_a} \right) \right) \nonumber \\
&+ \sum_{c,d \neq a,b} \frac{(v_c-v_a)(v_d-v_a)}{(v_b-v_a)(v_d-v_c)} \left( X^a_cX^b_d - X^a_dX^b_c \right).
\end{align}
In the GB, we must solve $2M$ linear equations, each with cost $\mathcal{O}(M^3)$. While the sums to be computed are less clean they are no more difficult to compute numerically. Their construction scales like $\mathcal{O}(M^4)$.

\section{Conclusions}
In this contribution, we report optimal expressions for the 1- and 2-RDMs of the RG states in both the physical and Gaudin bases. All RDM expressions are evaluated from solutions to sets of linear equations, which all share the same matrix. For large systems this matrix will naturally become sparse as the off-diagonal entries quickly go to zero. Occasional numerical instability may arise from Laguerre's method failing to produce rapidities from EBV, in which case RDM expressions directly in terms of the EBV would be more robust, but would not beat the scaling. Practical expressions for the scalar products in terms of EBV are known, though 2-RDM expressions are not. We thus consider the problem of finding the numerically cheapest method to evaluate our objective function solved. In following contributions we will consider transition density matrices between RG states, which is the next clear step towards a perturbation theory.

\section{Data Availability}
The data that support the findings of this study are available from the corresponding author upon reasonable request.

\section{Acknowledgements}
We gratefully acknowledge support from the Natural Sciences and Engineering Research Council of Canada.

\appendix
\section{Second Derivatives of Richardson's equations}
The second derivatives of Richardson's equations required to calculate the analytic gradient are:

\begin{align}
0 &= 2 \left( \left(
\sum_i \frac{1}{(u_a-\varepsilon_i)^3} + \sum_{b \neq a} \frac{2}{(u_b-u_a)^3} \right) \frac{\partial u_a}{\partial \varepsilon_k} - \sum_{b \neq a} \frac{2}{(u_b-u_a)^3}\frac{\partial u_b}{\partial \varepsilon_k} -\frac{1}{(u_a-\varepsilon_k)^3}
\right) \frac{\partial u_a}{\partial \varepsilon_k} \nonumber \\
&+ \left( \sum_{ b \neq a} \frac{2}{(u_b-u_a)^2} - \sum_i \frac{1}{(u_a-\varepsilon_i)^2} \right)\frac{\partial^2 u_a}{\partial \varepsilon_k^2}
- \sum_{b\neq a}\frac{2}{(u_b-u_a)^2} \frac{\partial^2 u_b}{\partial \varepsilon_k^2}
+ \sum_{b\neq a} \frac{4}{(u_b-u_a)^3} \frac{\partial u_b}{\partial \varepsilon_k} \frac{\partial u_b}{\partial \varepsilon_k} \nonumber \\
&- \sum_{b\neq a} \frac{4}{(u_b-u_a)^3} \frac{\partial u_a}{\partial \varepsilon_k} \frac{\partial u_b}{\partial \varepsilon_k} 
- \frac{2}{(u_a-\varepsilon_k)^3}\frac{\partial u_a}{\partial \varepsilon_k} + \frac{2}{(u_a-\varepsilon_k)^3}
\end{align}

\begin{align}
0 &= 2 \left( \left(
\sum_i \frac{1}{(u_a-\varepsilon_i)^3} + \sum_{b \neq a} \frac{2}{(u_b-u_a)^3} \right) \frac{\partial u_a}{\partial \varepsilon_l} - \sum_{b \neq a} \frac{2}{(u_b-u_a)^3}\frac{\partial u_b}{\partial \varepsilon_l} -\frac{1}{(u_a-\varepsilon_l)^3}
\right) \frac{\partial u_a}{\partial \varepsilon_k} \nonumber \\
&+ \left( \sum_{ b \neq a} \frac{2}{(u_b-u_a)^2} - \sum_i \frac{1}{(u_a-\varepsilon_i)^2} \right)\frac{\partial^2 u_a}{\partial \varepsilon_l \partial \varepsilon_k}
- \sum_{b\neq a}\frac{2}{(u_b-u_a)^2} \frac{\partial^2 u_b}{\partial \varepsilon_l \partial \varepsilon_k}
+ \sum_{b\neq a} \frac{4}{(u_b-u_a)^3} \frac{\partial u_b}{\partial \varepsilon_l} \frac{\partial u_b}{\partial \varepsilon_k} \nonumber \\
&- \sum_{b\neq a} \frac{4}{(u_b-u_a)^3} \frac{\partial u_a}{\partial \varepsilon_l} \frac{\partial u_b}{\partial \varepsilon_k} 
- \frac{2}{(u_a-\varepsilon_k)^3}\frac{\partial u_a}{\partial \varepsilon_l}
\end{align}

\begin{align}
0 &= 2 \left( \left(
\sum_i \frac{1}{(u_a-\varepsilon_i)^3} + \sum_{b \neq a} \frac{2}{(u_b-u_a)^3} \right) \frac{\partial u_a}{\partial g} - \sum_{b \neq a} \frac{2}{(u_b-u_a)^3}\frac{\partial u_b}{\partial g }
\right) \frac{\partial u_a}{\partial \varepsilon_k} \nonumber \\
&+ \left( \sum_{ b \neq a} \frac{2}{(u_b-u_a)^2} - \sum_i \frac{1}{(u_a-\varepsilon_i)^2} \right)\frac{\partial^2 u_a}{\partial g \partial \varepsilon_k}
- \sum_{b\neq a}\frac{2}{(u_b-u_a)^2} \frac{\partial^2 u_b}{\partial g \partial \varepsilon_k}
+ \sum_{b\neq a} \frac{4}{(u_b-u_a)^3} \frac{\partial u_b}{\partial g} \frac{\partial u_b}{\partial \varepsilon_k} \nonumber \\
&- \sum_{b\neq a} \frac{4}{(u_b-u_a)^3} \frac{\partial u_a}{\partial g} \frac{\partial u_b}{\partial \varepsilon_k} 
- \frac{2}{(u_a-\varepsilon_k)^3}\frac{\partial u_a}{\partial g}
\end{align}

\bibliography{BAbasis}

\bibliographystyle{unsrt}

\end{document}